\documentclass[12pt]{article} 
\usepackage{graphicx}
\usepackage{amssymb}
\usepackage{amsmath}
\usepackage{multirow}
\usepackage{float} 
\usepackage{cite}
\usepackage{soulutf8}
\usepackage{titlesec}
\usepackage[singlelinecheck=false]{caption}
\usepackage{lscape}
\usepackage{array}
\usepackage{enumerate}
\usepackage{dsfont}
\usepackage{algorithm}
\usepackage{algorithmicx}
\usepackage{algpseudocode}
\usepackage{footmisc}
\usepackage{pdflscape}
\usepackage{color}
\usepackage{pstricks}
\usepackage{slashbox}

\titleformat{\section}[runin]{\normalfont\bfseries}{\thesection}{1em}{}[]
\titleformat{\subsection}[runin]{\normalfont}{\thesubsection}{1em}{\so}[]
\makeatletter\renewcommand{\@biblabel}[1]{#1.}\makeatother
\def\thesection{\indent \arabic{section}.}
\def\thesubsection{\indent \arabic{section}.\arabic{subsection}.}

\def\DiscrIntr#1#2{#1, \dots, #2}

\begin{document}
\noindent
QUANTUM PHYSICS\\
UDC 51-71, 515.173

\begin{center} 
\textbf{Space-time structure in the microcosm and its relation to the properties of elementary particles} \\
\vspace{4mm}
{\copyright  2020\;\;
Popov N.N.$^{*}$ \\
$^{*}\;$Federal Research Center ``Computer Science and Control'' of Russian Academy of Sciences, Moscow \\
e-mail: nnpopov@mail.ru \\ 
} 
\end{center}

{\footnotesize
The relations between the hidden symmetries of the six-dimensional pseudo-Euclidean space with signature $(+\; +\; +\; -\; -\; -)$ and the conserved quantum characteristics of elementary particles is established. 
The hidden symmetries are brought out by the various forms of representation of the pseudo-Euclidean space metric with the aid of spinors and hyperbolic complex numbers. 
Using the emerging hidden symmetry groups one can disclose such conserved quantum characteristics as spin, isospin, electric and baryon charges, hypercharge, color and flavor. 
One can also predict the exact number of such conserved quantum characteristics of quarks as color and flavor. 

}

\section*{Introduction.}

In the theory of grand unification based on the structural group $SU(3) \otimes $ $SU(2) \otimes$ $U(1)$ the mathematical number of space dimensions is equal to $11$. 
In this case four dimensions refer to real physical space-time, whereas the remaining seven correspond to some abstract space \cite{Witten81}, within the framework of which one can introduce such quantum characteristics of elementary particles as isospin, hypercharge, colour, flavour, etc. 
The number of dimension of that additional abstract space may increase as ever new conserved quantum characteristics of elementary particles are discovered. 
This scheme of development of the theory is rather simple, however one cannot get reed of the feeling that it is artificial. 

The concept put forward in the present paper can be presented in the following statements. 

1. The conserved quantum characteristics of elementary particles must be related to the geometrical properties of the real physical space-time; 

2. The four-dimensional pseudo-Riemannian space lacks the necessary set of geometrical properties to allow such a relationship to be established. 

3. One should revise current views of the properties of the real physical space-time in the microcosm and find such a structure of the manifold that the symmetry groups of its typical tangential layer could generate the conserved quantum characteristics of the elementary particles. 

4. The total number of permissible symmetry groups of the real space-time in the microcosm should coincide with the total number of the conserved quantum characteristics of elementary particles. 

As a candidate to the role of real physical space-time in the microcosm we will consider the six-dimensional pseudo-Riemannian space of signature $(+\; +\; +\; -\; -\; -)$ with the typical tangential layer in the form of pseudo-Euclidean space. 
It is assumed the in the six-dimensional space the world time flows along a certain axis in the three-dimensional time subspace. 
The presence of certain time axis leads to violation of spherical symmetry in the three-dimensional time subspace. 
As it will be shown below, it leads to violation of some conservation laws. 
However, in studying of physical processes in very small time intervals of the order $10^{-20}$s and less the three-dimensional time subspace may be regarded as isotropic with fair accuracy. 
It it this region where the possible space symmetry groups will be sought for. 

Such groups are brought out by various form of representations of the pseudo-Euclidean metric of the six-dimensional space. 
First of all this refers to the spinor representation form \cite{Popov18_1} and also to the metric representation usin hyperbolic complex numbers \cite{Popov18_1, Popov18_2}. 
It is shown here that the spinor metric representation form leads to finding such conservation laws for the quantum characteristics of elementary particles as spin, isospin, electric and baryon charges, hypercharge. 
It its turn the metric representation using the hyperbolic complex numbers leads to finding hyperbolic groups of unitary symmetry, which leave the metric of the six-dimensional pseudo-Euclidean space-time invariant and which generate such conserved quark characteristics as colour and flavour. 
In the case of quark model, using the hyperbolic groups of unitary symmetry allows to predict that there exists only three colours and strictly six quark flavours if the spherical symmetry is violated. 

\section{Pseudo-Euclidean space $\mathbb{E}_{3,3}$ as an image of the spinor space.} 

Consider the pseudo-Euclidean space $\mathbb{E}_{3,3}$ and let $\eta_{ij}$ designate the metric there, i.e.
\begin{equation}
\eta_{ij} = \begin{cases}
0\;,\;\; i \neq j \;, \\
+1\;,\;\; i=j=1,2,3 \;,\\
-1\;,\;\; i=j=4,5,6 \;.\\
\end{cases}
\end{equation}
The squared interval in $\mathbb{E}_{3,3}$ is given by
\begin{equation} \label{eq_sqinterval}
s^2 = \eta_{kl} x^k x^l \;,\;\; k,l = \DiscrIntr{1}{6} \;,
\end{equation}
where $\vec{x} = (x_1,\dots,x_6)$ is a vector in $\mathbb{E}_{3,3}$, and summation is meant by same upper and lower indices. 
The group of proper motions of metric \eqref{eq_sqinterval} in $\mathbb{E}_{3,3}$ is given by a group of proper pseudo-Euclidean rotations $SO(3,3)$.

Let us now introduce the four-dimensional complex space $\mathbb{C}^4$, whose elements are the four-component complex vectors, called spinors $\vec{\xi} = (\xi_1,\dots,\xi_4)$, and the space $\mathbb{C}^4$ itself is referred to as spinor space \cite{Rashevsky55}. 
Now denote $t^1 = x^4$, $t^2 = x^5$, $t^3 = x^6$. 

For any vector $\vec{x} \in \mathbb{E}_{3,3}$ one can find such a spinor  $\vec{\xi} \in \mathbb{C}^4$, that the following will hold:
\begin{equation} \label{eq_vecrefs}
\begin{split}
t^1 = \xi^1 \dot{\xi}^2 + \xi^2 \dot{\xi}^1 \;,&\;\; 
t^2 = \frac{\xi^1 \dot{\xi}^2 - \dot{\xi}^1 \xi^2}{i} \;,\;\;
t^3 = \xi^1 \dot{\xi}^1 - \xi^2 \dot{\xi}^2 \;,\\
x^1 = \xi^3 \dot{\xi}^4 + \dot{\xi}^3 \xi^4 \;,&\;\;
x^2 = \frac{\xi^3 \dot{\xi}^4 - \dot{\xi}^3 \xi^4}{i} \;,\;\;
x^3 = \xi^3 \dot{\xi}^3 - \xi^4 \dot{\xi}^4 \;,
\end{split}
\end{equation}
where $\dot{\xi}^{\mu}$ is the complex-conjugated to ${\xi}^{\mu}$ spinor component, and $i$ is the imaginary unit.
Formulas \eqref{eq_vecrefs} may be rewritten in more elegant form. 
To do so, consider the complex matrix algebra $M(4,\mathbb{C})$ in spinor space $\mathbb{C}^4$. 
In this algebra let us choose the matrices
\begin{equation} \label{eq_algebramatrices}
\widehat{\sigma}^p = \left\lbrace \begin{array}{l}
\;\;\left( \begin{array}{cc}\sigma^p&0\\0&0\end{array}\right)\;,\;\; p=1,2,3,\\
\left( \begin{array}{cc}0&0\\0&\sigma^{p-3}\end{array}\right)\;,\;\; p=4,5,6, 
\end{array} \right.
\end{equation}
where
\begin{equation} \label{eq_paulimatrices}
\sigma^1 = \left( \begin{array}{cc} 0& 1 \\ 1& 0 \end{array} \right)\;,\;\;
\sigma^2 = \left( \begin{array}{cc} 0& -i\\ i& 0 \end{array} \right)\;,\;\;
\sigma^3 = \left( \begin{array}{cc} 1& 0 \\ 0& -1 \end{array} \right)\;-
\end{equation}
are Pauli matrices. 
The matrices \eqref{eq_algebramatrices} form the six-dimensional basis in the subalgebra $L$ of the algebra $M(4,\mathbb{C})$. 
The following commutation relationships take place: 
\begin{equation} \label{eq_commutations}
\widehat{\sigma}^k \widehat{\sigma}^l - \widehat{\sigma}^l \widehat{\sigma}^k = 2i \widehat{\sigma} ^m \varepsilon_{klm} \;,
\end{equation}
where $k,l,m = \DiscrIntr{1}{6}$ , $\varepsilon_{klm}$ is the completely antisymmetric tensor, product $\widehat{\sigma}^k \widehat{\sigma}^l$ is zero if indices of the pair belong to different triplets ($k = 1, 2, 3$; $l = 4, 5, 6$), 
\begin{equation} \label{eq_anticommutations}
\widehat{\sigma}^k \widehat{\sigma}^l + \widehat{\sigma}^l \widehat{\sigma}^k = 2\delta^{kl} p^m \;,
\end{equation}
where $p^m$ is the two-dimensional orthogonal projector in the spinor space  $\mathbb{C}^4$. 
At that, if $k, l = 1, 2, 3$ then $m = 2$, if $k, l = 4, 5, 6$ then $m = 1$, $p^1(\mathbb{C}^4) = \{ \vec{\xi} \in \mathbb{C}^4; \vec{\xi} = (\xi^1, \xi^2, 0, 0)\}$, $p^2(\mathbb{C}^4) = \{\vec{\xi} \in \mathbb{C}^4; \vec{\xi} = (0, 0, \xi^3, \xi^4)\}$. 

According to the the relationships \eqref{eq_paulimatrices}, \eqref{eq_commutations}, \eqref{eq_anticommutations} the Lie algebra \eqref{eq_algebramatrices} is reducible. 
To each pair $(\widehat{\sigma}^m, \vec{\xi}) \in M(4,\mathbb{C}) \otimes \mathbb{C}^4$ there is a corresponding $m$-th coordinate of the vector $\vec{x} \in \mathbb{E}_{3,3}$ according to the formula
\begin{equation} \label{eq_scalar}
x^m = \langle \vec{\xi}, \widehat{\sigma}^m \vec{\xi} \rangle\;,\;\; m = \DiscrIntr{1}{6}\;,
\end{equation}
where $\langle \cdot , \cdot \rangle$ is the scalar product in $\mathbb{C}^4$, given by $\langle \vec{\xi}, \vec{\eta} \rangle = \delta_{\nu\mu}\dot{\xi}^\nu \eta^\mu$. 

One can readily notice that the representations \eqref{eq_vecrefs} and \eqref{eq_scalar} are equivalent. 
Thus the material coordinates of the vectors of pseudo-Euclidean space  $\mathbb{E}_{3,3}$ can be represented as the average values of the Hermitian operators of the form \eqref{eq_algebramatrices} upon spinors in $\mathbb{C}^4$ space. 

\section{Hidden groups of proper motions of the metric.}

From formula \eqref{eq_scalar} it follows that, if an arbitrary vector $\vec{x} \in \mathbb{E}_{3,3}$ with coordinates $(x^1, \dots, x^6)$ is set, then in general case there exists a pair $(\widehat{\sigma}_m, \vec{\xi}) \in M(4,\mathbb{C}) \otimes \mathbb{C}^4$, determinable to within the unitary equivalence with respect to the group $SU(4)$, and the relation \eqref{eq_scalar} holds. 
This means that the pseudo-Euclidean metric \eqref{eq_sqinterval} in spinor space $\mathbb{C}^4$ is invariant relative to the action of the group $SU(4)$, which may be regarded as a hidden of proper motions of this metric. 
Before studying the relation of the group $SU(4)$ with the conserved quantum characteristics of elementary particles we will consider simpler groups, leading us to conservation laws of such characteristics as spin, ``weak'' isospin, electric charge, and ``weak'' hypercharge. 

Consider the two-parametric group of unitary transformations $U(1) \oplus U(1)$, which is represented in $M(4,\mathbb{C})$ as unitary matrices of the following kind: 
\begin{equation} \label{eq_matrixfour}
\left( \begin{array}{cccc}
e^{i\varphi} & 0 & 0 & 0 \\ 0 & e^{i\varphi} & 0 & 0 \\
0 & 0 & e^{i\psi} & 0 \\ 0 & 0 & 0 & e^{i\psi}
\end{array} \right) \;,
\end{equation}
and which operates in $\mathbb{C}^4$. 
The transformations of the group $U(1) \oplus U(1)$ leave the right-hand sides of the relations \eqref{eq_vecrefs} invariant, i.e. the coordinates of the vectors in $\mathbb{E}_{3,3}$ remain unchanged under such transformations. 
Therefore, the metric \eqref{eq_sqinterval} itself of the space  $\mathbb{E}_{3,3}$ remains invariant. 
We will say that the transformations of the kind \eqref{eq_matrixfour} of the group $U(1) \oplus U(1)$ represent hidden motions of the metric \eqref{eq_sqinterval}. 
Group $U(1) \oplus U(1)$ generates two conservation laws. 
The first law generated by the operator $\frac{1}{i}\frac{\partial}{\partial \varphi}$ will be interpreted as the law of conservation of ``weak'' hypercharge. 
The second law, induced by the generator of the group $\frac{1}{i}\frac{\partial}{\partial \psi}$ will be interpreted as the law of conservation of the electric charge. 
The appearance of the ``weak'' hypercharge conservation law is related to the hidden symmetries in the three-dimensional time subspace, which for the sake of brevity will be referred to as isospace. 

Let us now proceed to the consideration of the more complicated unitary group. 
The representation of the unitary group with the Lie algebra, which is determined by the generators \eqref{eq_algebramatrices}, and which operates in the space $\mathbb{C}^4$, is quite reducible and may be expressed in the form of the direct sum of irreducible representations $SU(2) \oplus SU(2)$. 
Each of the irreducible representations corresponds to the group $SU(2)$ of unitary unimodular matrices $U$ of dimension $2$, i.e. $U^{+}U = 1$, $\det U = 1$. 
In case of the first irreducible representation such matrices may be represented in the form $U = e^{i\sigma_k a_k}$, $k = 1, 2, 3$, where $\sigma_k$ are the Hermitian Pauli matrices and $a_k$ are arbitrary real numbers. 
These matrices implement the identical representation of dimension $2$ in the two-dimensional isospin space $p^1(\mathbb{C}^4)$ with elements $\left( \begin{array}{c} \xi_1 \\ \xi_2 \end{array} \right)$ put over the two basis spinors $\left( \begin{array}{c} 1 \\ 0 \end{array} \right)$ and $\left( \begin{array}{c} 0 \\ 1 \end{array} \right)$. 
In the case of the second irreducible representation we will obtain the similar group of unitary unimodular matrices $U = e^{\sigma_l a_l}$, $l = 4, 5, 6$, where the same Pauli matrices $\sigma_l$ appear as generators according to the relation \eqref{eq_algebramatrices}. 
These matrices realize a group of dimension $2$ in the two-dimensional spinor space $p^2(\mathbb{C}^4)$ with elements $\left( \begin{array}{c} \xi_3 \\ \xi_4 \end{array} \right)$ put over two basis spinors. 
Thus the isospace is the spin space related to the three-dimensional time subspace. 
Therefore, such an important characteristic of elementary particle as weak isospin has a pure geometric nature and its conservation law is related to the invariance of the metric of the six-dimensional space $\mathbb{E}_{3,3}$ with respect to the group of rotations in the three-dimensional time subspace.

We note that the introduced laws of conservation of the ``weak'' hypercharge and isospin may be violated in the case of weak interactions but are not violated in the case of strong interactions. 

\section{Causes of violation of conservation laws of the hypercharge and isospin.}

Within the common formalism there is no explanation of the causes of violation of the conservation laws for hypercharge and isospin in weak interactions. 
In the proposed approach such a phenomenon is given rather simple explanation. 
According to the result obtained above the law of conservation of isospin and hypercharge has appeared due to the existence of spherical symmetry in the three-dimensional time subspace. 
If that space would always remained isotropic, the laws of conservation of isospin and hypercharge would be rigorous. 
However if the time axis along which the world time is counted has a preferable direction in the three-dimensional time subspace, then the spherical symmetry is violated and the above conservation laws are violated as well. 
Now it only remains to understand why these law are violated in the weak interactions only, but hold in the strong interactions. 

The point is that over very small time intervals the three-dimensional time subspace may be regarded as isotropic i.e. possesses spherical symmetry. 
String interactions run over a time of order $10^{-24}$s. 
During such small intervals the time subspace remains spherically symmetric and hence the conservation laws of hypercharge and isospin stay valid. 
In the case of weak interactions which run much slower during time intervals of $10^{-9}$s it is no longer possible to neglect the existence of the preferable time axis in such time intervals which leads to violation of spherical symmetry in the three-dimensional time subspace and consequently to violation of mentioned conservation laws. 

We note that the existence of the preferable time axis though leading to the violation of the spherical symmetry, nevertheless keeps the axial symmetry in the three-dimensional time subspace, which suggests the possibility of the existence of conservation laws related to the axial symmetry. 

\section{Group $SU(4)$ and the conserved quantum characteristics it generates.}

It has been shown above that the group $SU(4)$ leaves the metric of space $\mathbb{E}_{3,3}$ invariant. 
Let us now proceed to study in more detail the properties of the group $SU(4)$ and its Lie algebra for the necessity of giving their physical interpretations. 
The generators $\lambda_i$, $i = \DiscrIntr{1}{15}$ of Lie algebra may be represented as $15$ traceless Hermitian four-dimensional matrices:
\begin{equation} \label{eq_matricesfour}
{\footnotesize
\begin{split}
& \lambda_1 = \left( \begin{array}{cccc}
0&1&0&0\\1&0&0&0\\0&0&0&0\\0&0&0&0 \end{array} \right) ,\;
 \lambda_2 = \left( \begin{array}{cccc}
0&-i&0&0\\i&0&0&0\\0&0&0&0\\0&0&0&0 \end{array} \right) ,\;
 \lambda_3 = \left( \begin{array}{cccc}
1&0&0&0\\0&-1&0&0\\0&0&0&0\\0&0&0&0 \end{array} \right) ,\\
& \lambda_4 = \left( \begin{array}{cccc}
0&0&1&0\\0&0&0&0\\1&0&0&0\\0&0&0&0 \end{array} \right) ,\;
 \lambda_5 = \left( \begin{array}{cccc}
0&0&-i&0\\0&0&0&0\\i&0&0&0\\0&0&0&0 \end{array} \right) ,\;
 \lambda_6 = \left( \begin{array}{cccc}
0&0&0&0\\0&0&1&0\\0&1&0&0\\0&0&0&0 \end{array} \right) ,\\
& \lambda_7 = \left( \begin{array}{cccc}
0&0&0&0\\0&0&-i&0\\0&i&0&0\\0&0&0&0 \end{array} \right) ,\;
 \lambda_8 = \frac{1}{\sqrt{3}} \left( \begin{array}{cccc}
1&0&0&0\\0&1&0&0\\0&0&-2&0\\0&0&0&0 \end{array} \right) ,\;
 \lambda_9 = \left( \begin{array}{cccc}
0&0&0&1\\0&0&0&0\\0&0&0&0\\1&0&0&0 \end{array} \right) ,\\
& \lambda_{10} = \left( \begin{array}{cccc}
0&0&0&-i\\0&0&0&0\\0&0&0&0\\i&0&0&0 \end{array} \right) ,\;
 \lambda_{11} = \left( \begin{array}{cccc}
0&0&0&0\\0&0&0&1\\0&0&0&0\\0&1&0&0 \end{array} \right) ,\;
 \lambda_{12} = \left( \begin{array}{cccc}
0&0&0&0\\0&0&0&-i\\0&0&0&0\\0&i&0&0 \end{array} \right) ,\\ 
& \lambda_{13} = \left( \begin{array}{cccc}
0&0&0&0\\0&0&0&0\\0&0&0&1\\0&0&1&0 \end{array} \right) ,\;
 \lambda_{14} = \left( \begin{array}{cccc}
0&0&0&0\\0&0&0&0\\0&0&0&-i\\0&0&i&0 \end{array} \right) ,\;
 \lambda_{15} = \frac{1}{\sqrt{6}} \left( \begin{array}{cccc}
1&0&0&0\\0&1&0&0\\0&0&1&0\\0&0&0&-3 \end{array} \right) .
\end{split} }
\end{equation}
We note that Lie algebra for the group $SU(4)$ contains the Gell-Mann subalgebra for the group $SU(3)$, assigned by the generators $\lambda_1,\dots,\lambda_8$, as well as Pauli subalgebra for the group $SU(2)$, assigned by generators $\lambda_1,\dots,\lambda_3$. 
Let us introduce the following designations: 
\begin{equation} \label{eq_operators}
\begin{split}
& F_i = \frac{1}{2} \lambda_i \;,\;\; i=\DiscrIntr{1}{15} \;, \\
& I_\pm = F_1 \pm i F_2 \;,\;\; I_3 = F_3 \;, \\
& V_\pm = F_4 \pm i F_5 \;,\;\; U_\pm = F_6 \pm i F_7 \;,\;\; Y = \frac{2}{\sqrt{3}}F_8 \\
& N_\pm = F_9 \pm i F_{10} \;,\;\; M_\pm = F_{11} \pm i F_{12} \;, \;\; W_\pm = F_{13} \pm i F_{14} \;,\;\; B = \frac{4}{\sqrt{6}} F_{15} \;,
\end{split}
\end{equation}
where $I_\pm$ are the raising and lowering operators for isospin projection to rime axis, $Y$ if the Hermitian hypercharge operator, $B$ is the Hermitian baryon charge operator. 
Among the operators given by \eqref{eq_matricesfour} there is no one for the electrical charge $Q$. 
It can be assigned by the following Hermitian traceless matrix
\begin{equation}
Q = \frac{1}{3} \left( \begin{array}{cccc} 
2&0&0&0\\0&-1&0&0\\0&0&-1&0\\0&0&0&0 \end{array} \right) \;.
\end{equation}
Then the relation
\begin{equation} \label{eq_nisidzima}
Q= I_3+ \frac{Y}{2} \;,
\end{equation}
takes place, which was discovered phenomenologically by Gell-Mann \cite{Gelman53} and Nishijima \cite{Nishijima53}. 

Hermitian operators $I_3$, $Y$, $Q$, $B$ satisfy the following commutation relations:
\begin{equation} \label{eq_commuts}
[I_3, Y] = [I_3, Q] = [I_3, B] = [Y, Q] = [Y, B] = [Q, B] = 0 \;,
\end{equation}
i.e. all these four operators commutate. 
This means that the physical characteristics of elementary particles given by eigen values of these operators are simultaneously observable. 
Let us now find the expressions to describe the relationship between the commutating Hermitian operators. 
We will write out the commutators for the operators of creation and annihilation given by \eqref{eq_operators}: 
\begin{equation} \label{eq_birthdeath}
\begin{split}
& [I_+,I_-]= 2 I_3 \;,\;\; [V_+,V_-]= I_3+ \frac{3}{2}Y\;,\;\; [U_+,U_-]= -I_3+ \frac{3}{2}Y \;\\
& [N_+,N_-]= I_3+ \frac{1}{2}Y+ B\;,\;\; [M_+,M_-]= -I_3+ \frac{1}{2}Y+ B\;,\\ 
& [W_+, W_-] = -Y+B\;.
\end{split}
\end{equation}
Operator $Q$ may be used also, for instance $[V_+,V_-]=3Q-2I_3$. 

Let us tabulate the commutators, which are useful for constructing finite-dimensional representations of $SU(4)$ group. 
First component is given by row, second is assigned by column. 
Commutators of $I_3$, $Y$, $Q$, $B$ with the rest ones are presented in the table \ref{tab_commut}. 
\begin{table}[h] 
\caption{Commutators of operators \eqref{eq_operators}} 
\label{tab_commut}
 \centering
 \begin{tabular}{|l|c|c|c|c|c|c|}
 \hline
  \backslashbox{$\langle a, \cdot \rangle$}{$\langle \cdot, b \rangle$} & $I_\pm$ & $V_\pm$ & $U_\pm$ & $N_\pm$ & $M_\pm$ & $W_\pm$ \\
 \hline
 \hline
 $I_3$ & $\pm I_\pm$ & $\pm\frac{1}{2}V_\pm$ & $\mp\frac{1}{2}U_\pm$ & $\pm\frac{1}{2}N_\pm$ & $\mp\frac{1}{2}M_\pm$ & 0 \\
 \hline
 $Y$   & 0 & $\pm V_\pm$ & $\pm U_\pm$ & $\pm\frac{1}{3}N_\pm$ & $\pm\frac{1}{3}M_\pm$ & $\mp\frac{2}{3}W_\pm$ \\
 \hline
 $Q$ & $\pm I_\pm$ & $\pm V_\pm$ & 0 & $\pm\frac{2}{3}N_\pm$ & $\mp\frac{1}{3}M_\pm$ & $\mp\frac{1}{3}W_\pm$ \\
 \hline
 $B$ & 0 & 0 & 0 & $\pm\frac{4}{3}N_\pm$ & $\pm\frac{4}{3}M_\pm$ & $\pm\frac{4}{3}W_\pm$ \\
 \hline
 \end{tabular}
\end{table}

From the commutation relations given above if follows that operators $I_+$, $V_+$, $U_-$, $N_+$, $M_-$ are raising, and $I_-$, $V_-$, $U_+$, $N_-$, $M_+$ are lowering eigenvalues of $I_3$. 
Operators $U_+$, $V_+$, $N_+$, $M_+$, $W_-$ are raising, and $U_-$, $V_-$, $N_-$, $M_-$, $W_+$ are lowering eigenvalues of $Y$. 
Operators $I_+$, $V_+$, $N_+$, $M_-$, $W_-$ are raising, and $I_-$, $V_-$, $N_-$, $M_+$, $W_+$ are lowering eigenvalues of $Q$. 
Operators $N_+$, $M_+$, $W_+$ are raising, $N_-$, $M_-$, $W_-$ are lowering, and $I_\pm$, $V_\pm$, $U_\pm$ keep unchanged the eigenvalues of $B$.

Commutators of the rest of operators are given in table \ref{tab_commut_ext}. 
\begin{table}[h] 
\caption{Commutators of operators \eqref{eq_operators}, continue} 
\label{tab_commut_ext}
 \centering
 \begin{tabular}{|l|c|c|c|c|c|c|c|c|c|c|}
 \hline
 {\footnotesize
  \backslashbox{$\langle a, \cdot \rangle$}{$\langle \cdot, b \rangle$} } &$V_+$ &$V_-$ &$U_+$ &$U_-$ &$N_+$ &$N_-$ &$M_+$ & $M_-$ &$W_+$ &$W_-$\\
 \hline
 \hline
 $I_+$ & 0 & $-U_-$ & $V_+$ & 0 & 0 & $-M_-$ & $N_+$ & 0 & 0 & 0 \\
 \hline
 $I_-$ & $U_+$ & 0 & 0 & $-V_-$ & $M_+$ & 0 & 0 & $-N_-$ & 0 & 0 \\
 \hline
 $V_+$ &0 &\eqref{eq_birthdeath} &0 &$I_+$ &0 & -$W_-$ & 0 & 0 & $N_+$ & 0\\
 \hline
 $V_-$ &\eqref{eq_birthdeath} &0 &$-I_-$ &0 &$W_+$ &0 & 0 & 0 & 0 & $-N_-$\\
 \hline
 $U_+$ &0 &$I_-$ &0 &\eqref{eq_birthdeath} &0 &0 &0 &$-W_-$ &$M_+$ &0\\
 \hline
 $U_-$ &$-I_+$ &0 &\eqref{eq_birthdeath} &0 &0 &0 &$W_+$ &0 &0 &$-M_-$\\
 \hline
 $N_+$ &0 &$-W_+$ &0 &0 &0 &\eqref{eq_birthdeath} & 0 & $I_+$ & 0 & $V_+$\\
 \hline
 $N_-$ &$W_-$ &0 &0 &0 &\eqref{eq_birthdeath} &0 &$-I_-$ &0 & $-V_-$ & 0 \\
 \hline
 $M_+$ &0 &0 &0 &$-W_+$ &0 &$I_-$ &0 &\eqref{eq_birthdeath} &0 & $U_+$ \\
 \hline
 $M_-$ &0 &0 &$W_-$ &0 &$-I_+$ &0 &\eqref{eq_birthdeath} &0 & $-U_-$ &0 \\
 \hline
 $W_+$ &$-N_+$ &0 &$-M_+$ &0 &0 &$V_-$ &0 &$U_-$ &0 &\eqref{eq_birthdeath}\\
 \hline
 $W_-$ &0 &$N_-$ &0 &$M_-$ &$-V_+$ &0 &$-U_+$ &0 &\eqref{eq_birthdeath} &0\\
 \hline
 \end{tabular}
\end{table}

Using the commutation relations from tables \ref{tab_commut} and \ref{tab_commut_ext} we will construct, as an example, the simplest irreducible finite-dimensional representation for the $SU(4)$ group. 
The states which are contained in the representations are characterized by a set of values of the quantum characteristics $(I_3, Y, Q, B)$. 
Transitions between various states are introduced with the aid of the raising and lowering operators given above. 

Let us introduce four orthogonal vectors 
\begin{equation} \label{eq_four_vectors}
\left(\begin{array}{c}1\\0\\0\\0\end{array}\right)\;,\;\;
\left(\begin{array}{c}0\\1\\0\\0\end{array}\right)\;,\;\;
\left(\begin{array}{c}0\\0\\1\\0\end{array}\right)\;,\;\;
\left(\begin{array}{c}0\\0\\0\\1\end{array}\right)\;
\end{equation}
in the four-dimensional space induced by the set of quantum characteristics $(I_3, Y, Q, B)$. 
These vectors are the eigenvectors of the Hermitian operators $I_3$, $Y$, $Q$, $B$, given by matrices of the type of \eqref{eq_operators} and with the following eigenvalues: $I_3$: $(\frac{1}{2}, -\frac{1}{2}, 0, 0)$, $Y$: $(\frac{1}{3}, \frac{1}{3}, -\frac{2}{3}, 0)$, $Q$: $(\frac{2}{3}, -\frac{1}{3}, -\frac{1}{3}, 0)$, $B$: $(\frac{1}{3}, \frac{1}{3}, \frac{1}{3}, -1)$. 
Let the state $\Psi_u=$ $\Psi_{\frac{1}{2}, \frac{1}{3}, \frac{2}{3}, \frac{1}{3}}$ is set. 
It is determined as a representation vector, which satisfies the following relations: 
\begin{equation} \label{eq_quak_u_rels}
\begin{split}
I_3 \Psi_u = &\frac{1}{2} \Psi_u \;,\\
Y \Psi_u = &\frac{1}{3} \Psi_u \;,\\
Q \Psi_u = &\frac{2}{3} \Psi_u \;,\\
B \Psi_u = &\frac{1}{3} \Psi_u \;.\\
\end{split}
\end{equation}
This state is an eigenvector of operators $I_3$, $Y$, $Q$, $B$ and corresponds to $u$ quark.
Subjecting this state to the operator $V_-$ we obtain new state $\Psi_s=$ $\Psi_{0, -\frac{2}{3}, -\frac{1}{3}, \frac{1}{3}}$, which corresponds to $s$ quark: 
\begin{equation}
\begin{split}
& I_3 V_- \Psi_{\frac{1}{2}, \frac{1}{3}, \frac{2}{3}, \frac{1}{3}} = (V_- I_3 - \frac{1}{2} V_-) \Psi_{\frac{1}{2}, \frac{1}{3}, \frac{2}{3}, \frac{1}{3}} = 0 \cdot V_- \Psi_{\frac{1}{2}, \frac{1}{3}, \frac{2}{3}, \frac{1}{3}} \;,\\
& Y V_- \Psi_{\frac{1}{2}, \frac{1}{3}, \frac{2}{3}, \frac{1}{3}} = (V_- Y - V_-) \Psi_{\frac{1}{2}, \frac{1}{3}, \frac{2}{3}, \frac{1}{3}} = -\frac{2}{3} V_- \Psi_{\frac{1}{2}, \frac{1}{3}, \frac{2}{3}, \frac{1}{3}} \;,\\
& Q V_- \Psi_{\frac{1}{2}, \frac{1}{3}, \frac{2}{3}, \frac{1}{3}} = (V_- Q - V_-) \Psi_{\frac{1}{2}, \frac{1}{3}, \frac{2}{3}, \frac{1}{3}} = -\frac{1}{3} V_- \Psi_{\frac{1}{2}, \frac{1}{3}, \frac{2}{3}, \frac{1}{3}} \;,\\
& B V_- \Psi_{\frac{1}{2}, \frac{1}{3}, \frac{2}{3}, \frac{1}{3}} = V_- B \Psi_{\frac{1}{2}, \frac{1}{3}, \frac{2}{3}, \frac{1}{3}} = \frac{1}{3} V_- \Psi_{\frac{1}{2}, \frac{1}{3}, \frac{2}{3}, \frac{1}{3}} \;.
\end{split}
\end{equation} 
This can be written briefly as
\begin{equation}
V_- \Psi_u = \Psi_s \;.
\end{equation}
The state with same set of characteristics corresponds to $b$ quark. 
It may be obtained by applying $N_- W_+$ operators subsequently to $\Psi_u$: 
\begin{equation}
N_- W_+ \Psi_u = \Psi_b \;.
\end{equation}
Next, applying $U_+$ operator to $\Psi_s$ state and using commutators from the table \ref{tab_commut_ext}, one can obtain a state $\Psi_d=$ $\Psi_{-\frac{1}{2}, \frac{1}{3}, -\frac{1}{3}, \frac{1}{3}}$, corresponding to $d$ quark: 
\begin{equation}
U_+ \Psi_s = \Psi_d \;.
\end{equation}
From $\Psi_d$ state using $I_+$ operator one may transfer to $\Psi_u$ state, or with the help of $M_-$ operator a state $\Psi_a=$ $\Psi_{0, 0, 0, -1}$ may be obtained, which corresponds to anti-baryon with the characteristics of $I_3 = 0$, $Y=0$, $Q=0$, $B=-1$: 
\begin{equation}
I_+ \Psi_d = \Psi_u\;,\;\; M_- \Psi_d = \Psi_a\;.
\end{equation} 
Four states $\Psi_u$, $\Psi_s$, $\Psi_d$, $\Psi_a$ establish a space of simplest irreducible finite-dimensional representation of $SU(4)$ group. 
In this space a state $\Psi_c=$ $\Psi_{0, \frac{4}{3}, \frac{2}{3}, \frac{1}{3}}$ also exists.
It corresponds to $c$ quark, and can be reached from $\Psi_d$ by $V_+$ operator or from $\Psi_u$ with the help of $U_+$ operator:
\begin{equation}
V_+ \Psi_d = \Psi_c\;,\;\; U_+ \Psi_u = \Psi_c\;.
\end{equation} 
The state with the same set of characteristics $(I_3, Y, Q, B)$, which is related to $t$ quark, can be obtained from $\Psi_d$ and $\Psi_u$ as 
\begin{equation}
N_+ W_- \Psi_d = \Psi_t\;,\;\; M_+ W_- \Psi_u = \Psi_t\;.
\end{equation} 

\begin{figure}[h]
\begin{pspicture}(0,0)(15,6)
{\large
 \rput(0,-0.5){
  \rput[B](7,0.5){{\bf s, b} $\left( 0, -\frac{2}{3}, -\frac{1}{3}, \frac{1}{3} \right)$}
  \rput[B](8.9,1.2){$V_-$}
  \psline{->}(8.9,1.9)(8,1)
  \rput[B](5.2,1.2){$U_-$}
  \psline{->}(5.1,1.9)(6,1)
  \rput[B](3.5,3.3){{\bf d} $\left( -\frac{1}{2}, \frac{1}{3}, -\frac{1}{3}, \frac{1}{3} \right)$}
  \rput[B](4.3,2.1){$U_+$}
  \psline{->}(4.9,2.1)(4,3)
  \rput[B](4.2,4.3){$V_-$}
  \psline{->}(4.9,4.7)(4,3.8)
  \rput[B](7,6.1){{\bf c, t} $\left( 0, \frac{4}{3}, \frac{2}{3}, \frac{1}{3} \right)$}
  \rput[B](5.2,5.3){$V_+$}
  \psline{->}(5.1,4.9)(6,5.8)
  \rput[B](8.9,5.3){$U_+$}
  \psline{->}(8.9,4.9)(8,5.8)
  \rput[B](10.5,3.3){{\bf u} $\left( \frac{1}{2}, \frac{1}{3}, \frac{2}{3}, \frac{1}{3} \right)$}
  \rput[B](9.9,4.3){$U_-$}
  \psline{->}(9.1,4.7)(10,3.8)
  \rput[B](9.9,2.2){$V_+$}
  \psline{->}(9.1,2.1)(10,3)
    \rput[B](8.1,3.7){$I_+$}
    \psline{->}(7.3,3.5)(8.7,3.5)
    \rput[B](6.5,3.7){$I_-$}
    \psline{->}(7.1,3.5)(5.7,3.5)
 }
}
\end{pspicture}
\caption{Diagram 2.}
	\label{fig_transitions}
\end{figure}
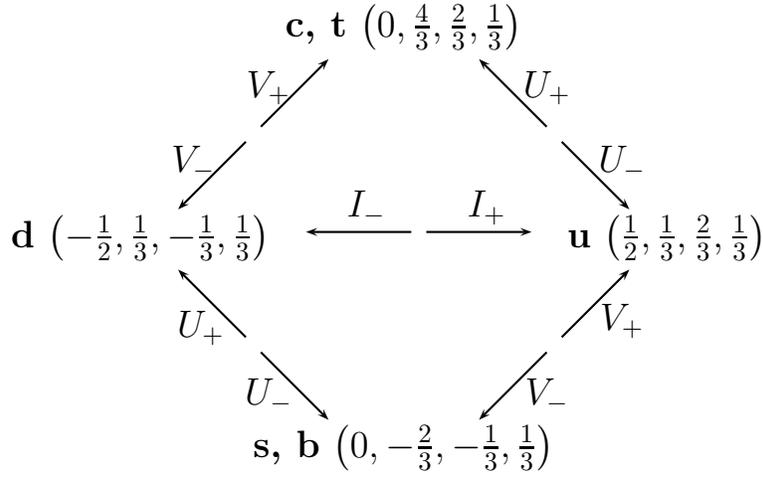

These results can be illustrated in the form of two transition diagrams (Fig.\ref{fig_transitions_prim}, \ref{fig_transitions}).

\begin{figure}[h]
\begin{pspicture}(0,0)(15,6)
{\large
 \rput(0,-0.5){
  \rput[B](7,0.5){{\bf s, b} $\left( 0, -\frac{2}{3}, -\frac{1}{3}, \frac{1}{3} \right)$}
  \rput[B](8.9,1.2){$V_-$}
  \psline{->}(8.9,1.9)(8,1)
  \rput[B](5.2,1.2){$U_-$}
  \psline{->}(5.1,1.9)(6,1)
  \rput[B](3.5,3.3){{\bf d} $\left( -\frac{1}{2}, \frac{1}{3}, -\frac{1}{3}, \frac{1}{3} \right)$}
  \rput[B](4.3,2.1){$U_+$}
  \psline{->}(4.9,2.1)(4,3)
  \rput[B](4.1,4.3){$M_+$}
  \psline{->}(4.9,4.7)(4,3.8)
  \rput[B](7,6.1){{\bf a} $\left( 0, 0, 0, -1 \right)$}
  \rput[B](5.1,5.3){$M_-$}
  \psline{->}(5.1,4.9)(6,5.8)
  \rput[B](8.9,5.3){$N_-$}
  \psline{->}(8.9,4.9)(8,5.8)
  \rput[B](10.5,3.3){{\bf u} $\left( \frac{1}{2}, \frac{1}{3}, \frac{2}{3}, \frac{1}{3} \right)$}
  \rput[B](9.9,4.3){$N_+$}
  \psline{->}(9.1,4.7)(10,3.8)
  \rput[B](9.9,2.2){$V_+$}
  \psline{->}(9.1,2.1)(10,3)
    \rput[B](8.1,3.7){$I_+$}
    \psline{->}(7.3,3.5)(8.7,3.5)
    \rput[B](6.5,3.7){$I_-$}
    \psline{->}(7.1,3.5)(5.7,3.5)
 }
}
\end{pspicture}
\caption{Diagram 1.}
	\label{fig_transitions_prim}
\end{figure}
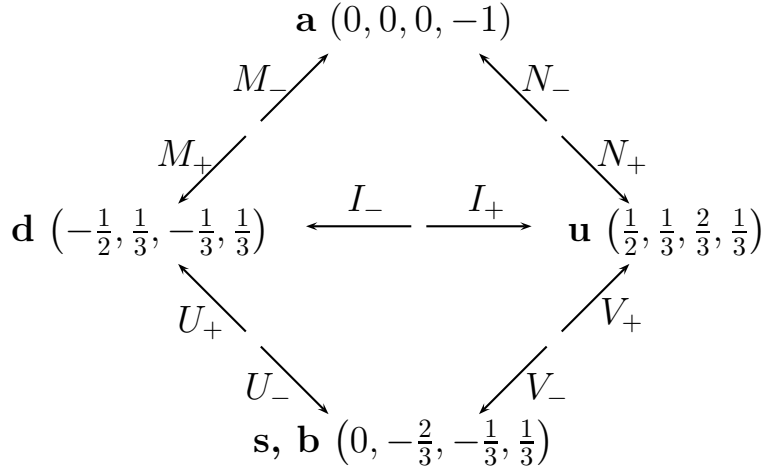

Diagram \ref{fig_transitions_prim} shows the simplest irreducible non-trivial representation of $SU(4)$ group, containing four quarks in three states and one anti-baryon with proper state $\Psi_{0, 0, 0, -1}$. 
Diagram \ref{fig_transitions} depicts simple non-trivial representation containing siz quarks in four states. 
At that states $\Psi_{\frac{1}{2}, \frac{1}{3}, \frac{2}{3}, \frac{1}{3}}$ and $\Psi_{-\frac{1}{2}, \frac{1}{3}, -\frac{1}{3}, \frac{1}{3}}$, correspond to quarks $u$ and $d$ respectively and appear once. 
States $\Psi_{0, -\frac{2}{3}, -\frac{1}{3}, \frac{1}{3}}$ and $\Psi_{0, \frac{4}{3}, \frac{2}{3}, \frac{1}{3}}$ correspond to quarks $(s, b)$ and $(c, t)$and appear two times each. 
In order to distinguish $s$ from $b$ and $c$ from $t$ it is necessary to have one more quantum characteristic. 
Such does not exist in the frame of $SU(4)$ group. 
The question about the total number of allowable quarks existing in the physical space also remains open. 
These questions are answered below. 

\section{Representation of the metric of $\mathbb{E}_{3,3}$ of space using hyperbolic complex numbers.}

Let us define the algebra $\mathbb{H}$ of hyperbolic complex numbers as a two-dimensional $R$-module with a pair of generatrixes $\{1, j\}$ and the following multiplication table
\begin{equation} \label{eq_multtable}
\begin{tabular}{ c|c|c|}
      & 1 & j \\
    \hline
    1 & 1 & j \\
    \hline
    j & j & 1 \\
    \hline
\end{tabular}
\end{equation}

The elements $h \in \mathbb{H}$ will be written in the form $h = 1x+jt$, where $x, t \in \mathbb{R}$, and $j$ is the imaginary unity in $\mathbb{H}$. 
Real numbers $\Re h = x$ and $\Im h = t$ are called real and imaginary parts of the hyperbolic complex number $h$ respectively. 
The involutive operation of complex conjugation is defined as $h = x+jt \rightarrow h = x-jt$. 
In a plane the algebra of hyperbolic complex numbers induces a two-dimensional pseudo-Euclidean geometry with metric $\eta_{ij} = \left( \begin{array}{cc} 1&0\\ 0&-1 \end{array} \right)$.
Consider now the $n$-dimensional space of hyperbolic complex numbers $\mathbb{H}^n$. 
Here we can introduce the scalar product of vectors $\langle \cdot , \cdot \rangle$. 
If $\vec{h} = (h^1, \dots, h^n)$, $\vec{g} = (g^1, \dots, g^n) \in \mathbb{H}^n$, then the scalar product is given by a bilinear form as: 
\begin{equation} \label{eq_scalar_product}
\langle \vec{h} , \vec{g} \rangle = h^1 \overline{g}^1 + \dots + h^n \overline{g}^n \;.
\end{equation}
The form \eqref{eq_scalar_product} is not positively defined. 
In case of $\mathbb{H}^3$ we have
\begin{equation} \label{eq_scalar_threedim}
\langle \vec{h}, \vec{h} \rangle = h^1 \overline{h}^1 + h^2 \overline{h}^2 + h^3 \overline{h}^3 \;.
\end{equation}
Taking into account that $h^k = x^k + jt^k, k = 1, 2, 3$ we obtain
\begin{equation} \label{eq_scalar_threedim_full}
\langle \vec{h}, \vec{h} \rangle = (x^1)^2 + (x^2)^2 + (x^3)^2 - (t^1)^2 - (t^2)^2 - (t^3)^2 \;,
\end{equation}
i.e. the scalar product of hyperbolic complex vectors from $\mathbb{H}^3$ assigns square bilinear form (pseudo-Euclidean metric) in $\mathbb{E}_{3,3}$ space. 
Now we consider some symmetry groups of the form \eqref{eq_scalar_threedim}.

\section{Hyperbolic groups of unitary symmetry and their representations.}

The metric of the six-dimensional pseudo-Euclidean space $\mathbb{E}_{3,3}$ is invariant relative to a number of hidden symmetry groups, which emerge as a result of representing the pseudo-Euclidean metric with the aid of hyperbolic complex numbers in $\mathbb{H}^3$ space according to \eqref{eq_scalar_threedim}.
Consider the unitary hyperbolic group $U(1,\mathbb{H}^3)$, which operates in $\mathbb{H}^3$ space. 
It is s three-parametric group of $H$-unitary matrices
\begin{equation} \label{eq_unitary_matrix}
U = \left( \begin{array}{ccc}
e^{j\varphi_1}&0&0\\ 0&e^{j\varphi_2}&0\\ 0&0&e^{j\varphi_3}
\end{array} \right) \;,
\end{equation}
which leave the bilinear form \eqref{eq_scalar_threedim} invariant.
The Hermitian-conjugated matrix 
\begin{equation}
U = \left( \begin{array}{ccc}
e^{-j\varphi_1}&0&0\\ 0&e^{-j\varphi_2}&0\\ 0&0&e^{-j\varphi_3}
\end{array} \right)
\end{equation}
is inverse of $U$, and $U U^{+} = 1$. 
Lie algebra of this group is commutative and its basis is formed as
\begin{equation}
e_1 = \left(\begin{array}{ccc}1&0&0\\0&0&0\\0&0&0\end{array}\right)\;,\;\;
e_2 = \left(\begin{array}{ccc}0&0&0\\0&1&0\\0&0&0\end{array}\right)\;,\;\;
e_3 = \left(\begin{array}{ccc}0&0&0\\0&0&0\\0&0&1\end{array}\right)\;.
\end{equation}

The identical representation \eqref{eq_unitary_matrix} of the group $U(1,\mathbb{H}^3)$ is reducible as a direct sum of irreducible representations, which operate in invariant single-dimension subspaces $\mathbb{H}$: 
\begin{equation}
U(1,\mathbb{H}^3) = U(1,\mathbb{H})\oplus U(1,\mathbb{H})\oplus U(1,\mathbb{H}) \;.
\end{equation}
Generators of this group induce three conservation laws. 
Running a little bit ahead we note that these conservation laws are associated with three color quantum characteristics of quarks. 
And the unitary transformations of the group $U(1,\mathbb{H})$ in $\mathbb{H}$ space correspond to Lorenz transformations in the pseudo-Euclidean space $\mathbb{E}_{3,3}$.

Consider now the hyperbolic group of unitary matrices $SU(2,\mathbb{H})$, which operates in the three-dimensional hyperbolic space $\mathbb{H}^3$. 
This group $SU(2,\mathbb{H})$ consists of hyperbolic matrices $U$ with dimension $2 \oplus 2$, which are unitary unimodular i.e. satisfy the conditions $U^{+}U = 1$, $| \det U| = 1$. 
Such a matrix may be represented as $U = e^{j\sigma_k a_k}$, $U^{+} = e^{-j\sigma_k a_k}$, where $\sigma_k$ are Hermitian traceless matrices, having the form of
\begin{equation} \label{eq_sigma_matrices}
\sigma_1 = \left( \begin{array}{cc} 0&1\\1&0 \end{array} \right) \;,\;\;
\sigma_2 = \left( \begin{array}{cc} 0&-j\\j&0 \end{array} \right) \;,\;\;
\sigma_3 = \left( \begin{array}{cc} 1&0\\0&-1 \end{array} \right) \;,\end{equation}
where $a_k$ are arbitrary real numbers. 

The matrices \eqref{eq_sigma_matrices} form a three-dimensional basis in the Lie algebra of $SU(2,\mathbb{H})$ group and differ from Pauli matrices only by replacing of imaginary unit $i$ with hyperbolic imaginary unit $j$. 
The basis elements \eqref{eq_sigma_matrices} of the Lie algebra satisfy the following commutation relations:
\begin{equation}
[\sigma_k, \sigma_l] = 2j\kappa_{klm}\sigma_m \;,
\end{equation}
where $\kappa_{klm}$ is a third-rank tensor with values $\kappa_{123} = 1$, $\kappa_{132} = 1$, $\kappa_{231} = 1$, $\kappa_{312} = -1$, $\kappa_{213} = -1$, $\kappa_{321} = -1$.

The structural constants of Lie algebra of the group $SU(2,\mathbb{H})$ coincide with that of group $SU(2)$ to within a sign.

From the components of six-dimensional vector $\vec{x} = (x^1, x^2, x^3, t^1, t^2, t^3) \in \mathbb{E}_{3,3}$ one can choose three so as to avoid all three being of the same type, i.e. triplets $(x^1, x^2, x^3)$ and $(t^1, t^2, t^3)$ are excluded. 
There are eighteen such triplets. 
They can be joined in pairs so as to have all six components in a pair. 
Fro instance for the triplet $(x^1, x^3, t^2)$ its pair will be $(x^2, t^1, t^3)$. 
Thus there are nine pairs. 

For each triplet of type $(x^k, x^l, t^m)$, which contains two spatial coordinates we assign a matrix
\begin{equation} \label{eq_matrix_for_triplets}
Y = \left( \begin{array}{cc} 
x^k & x^l-jt^m \\ x^l+jt^m & -x^k \end{array} \right) \;,
\end{equation}
and to its pair $(x^n, t^p, t^q)$, $n, p, q = 1, 2, 3$, $n \neq k, l$, $m \neq p, q$, we assign a matrix
\begin{equation} \label{eq_matrix_for_triplets_conj}
Y^C = \left( \begin{array}{cc}
t^p & t^q - jx^n \\ t^q+jx^n & -t^p
\end{array} \right) \;.
\end{equation}
The the following relation takes place: 
\begin{equation} \label{eq_matrix_for_triplets_det}
\det Y^C - \det Y = (x^1)^2+ (x^2)^2+ (x^3)^2- (t^1)^2- (t^2)^2- (t^3)^2 \;.
\end{equation}
For any hyperbolic unitary matrices $U_1, U_2 \in SU(2,\mathbb{H})$ the independent unitary transformations
\begin{equation}
Y^\prime = U_1^{+} Y U_1 \;,\;\; Y^{C\prime} = U_2^{+} Y^C U_2 
\end{equation}
leave the bilinear square form in the right side of \eqref{eq_matrix_for_triplets_det} invariant, by the equality
\begin{equation}
\det Y^{C\prime}- \det Y^\prime = \det Y^C - \det Y \;.
\end{equation}
Thus the unitary transformations over the pairs of matrices $Y$ and $Y^C$ from $SU(2,\mathbb{H})$ group correspond to the pseudo-orthogonal transformations in space $\mathbb{E}_{3,3}$, leaving the pseudo-Euclidean metric invariant. 

There are altogether nine groups of this kind each representation of such group expands into direct sum of two irreducible conjugated representations, which operate in three-dimensional subspaces of the six-dimensional space-time. 
To these eighteen representations of the hyperbolic unitary symmetry groups there should correspond eighteen conservation laws. 
Running ahead, we note that the conserved quantum characteristics may be interpreted as quark flavours. 
The fact that eighteen representations are paired means that quark flavours appear in pairs $(u, d)$, $(s, c)$, $(b, t)$ et.c. 

We note that the obtained result rests on the assumption that the time subspace is isotropic, i.e. there is no preferable time axis. 
However, if this is not so and such preferable time direction exists, let it be $t_1$, then the spherical symmetry in the time subspace is violated and only axial symmetry remains. 
In this case only three pairs are left: 
\begin{equation}
\begin{split}
&(t_1, x_1, x_2) \;\leftrightarrow\; (t_2, t_3, x_3) \;; \\ 
&(t_1, x_1, x_3) \;\leftrightarrow\; (t_2, t_3, x_2) \;; \\
&(t_1, x_2, x_3) \;\leftrightarrow\; (t_2, t_3, x_1) \;.
\end{split}
\end{equation}
To each pair there corresponds a pair of conjugated matrices \eqref{eq_matrix_for_triplets} and \eqref{eq_matrix_for_triplets_conj}, inducing six conservation laws for quark flavours. 

Let us introduce an operator
\begin{equation}
F = -\sigma_3 = \left( \begin{array}{cc} -1&0\\0&1 \end{array} \right)
\end{equation}
and consider the representation of the group $SU(2,\mathbb{H})$ in the six-dimensional space. 
Operators $\sigma_1$, $\sigma_2$, $\sigma_3$ \eqref{eq_sigma_matrices} and $F$ then take the form
\begin{equation}
{\footnotesize
\begin{split}
& \sigma_1 = \left( \begin{array}{ccc}
0&\hspace{-10pt}0&\hspace{-10pt}\boxed{\begin{array}{cc}0&1\\1&0\end{array}}\\
0&\hspace{-10pt}\boxed{\begin{array}{cc}0&1\\1&0\end{array}}&\hspace{-10pt}0\\
\boxed{\begin{array}{cc}0&1\\1&0\end{array}}&\hspace{-10pt}0&\hspace{-10pt}0
\end{array} \right) ,\;
  \sigma_2 = \left( \begin{array}{ccc}
0&\hspace{-10pt}0&\hspace{-10pt}\boxed{\begin{array}{cc}0&-j\\j&0\end{array}}\\
0&\hspace{-10pt}\boxed{\begin{array}{cc}0&-j\\j&0\end{array}}&\hspace{-10pt}0\\
\boxed{\begin{array}{cc}0&-j\\j&0\end{array}}&\hspace{-10pt}0&\hspace{-10pt}0
\end{array} \right),\\
& \sigma_3 = \left( \begin{array}{ccc}
\boxed{\begin{array}{cc}1&0\\0&-1\end{array}}&\hspace{-10pt}0&\hspace{-10pt}0\\
0&\hspace{-10pt}\boxed{\begin{array}{cc}1&0\\0&-1\end{array}}&\hspace{-10pt}0\\
0&\hspace{-10pt}0&\hspace{-10pt}\boxed{\begin{array}{cc}1&0\\0&-1\end{array}}
\end{array} \right) ,\;
  F = \left( \begin{array}{ccc}
\boxed{\begin{array}{cc}-1&0\\0&1\end{array}}&\hspace{-10pt}0&\hspace{-10pt}0\\
0&\hspace{-10pt}\boxed{\begin{array}{cc}-1&0\\0&1\end{array}}&\hspace{-10pt}0\\
0&\hspace{-10pt}0&\hspace{-10pt}\boxed{\begin{array}{cc}-1&0\\0&1\end{array}}
\end{array} \right) .
\end{split}}
\end{equation}
Six orthonormal vectors $(1, 0, \dots, 0)$, $(0, 1, \dots, 0)$, $\dots$ , $(0, \dots, 0, 1)$ describe six quark flavours designated as $d$, $u$, $s$, $c$, $b$, $t$. 
The representation of Hermitian operator $F$ in six-dimensional space assigns the flavour operator of six quarks. 
Six introduced orthonormal vectors are the eigenvectors of $F$. 
Eigenvalues of $F$ are $\pm 1$, flavours $u$, $c$, $t$ have eigenvalue of $+1$, and flavours $d$, $s$, $b$ have $-1$.
If we denote $\lambda_\alpha$ as an eigenvalue for flavour $\alpha$, then a simple relation exists between eigenvalues of $F$, baryon charges $B$ and electrical charges $Q_\alpha$: 
\begin{equation} \label{eq_simple_relation}
Q_\alpha = \frac{B+ \lambda_\alpha}{2} \;.
\end{equation}
Taking into account that the isospin projections $I_3$ of quarks $d$ and $u$ and eigenvalues $\lambda_d$ and $\lambda_u$ are formally related as
\begin{equation}
I_3(d) = \frac{\lambda_d}{2}\;,\;\; I_3(u) = \frac{\lambda_u}{2}\;,
\end{equation}
the formula \eqref{eq_simple_relation} may be rewritten as
\begin{equation} \label{eq_relation_next}
Q_\alpha = \begin{cases} 
I_3(\alpha)+\frac{B}{2}\;,\;\;\text{if }\alpha\text{ equals }d\text{ or }u \\ 
I_3(\alpha)+\frac{B+\lambda_\alpha}{2}\;,\;\;\text{if }\alpha\text{ equals }s, c, b, t
\end{cases}
\end{equation}
Comparing \eqref{eq_relation_next} and \eqref{eq_nisidzima}, one can see that there is a simple connection between baryon charges, flavours and hypercharges:

$Y_\alpha = B + \lambda_\alpha$, if $\alpha$ is one of $s$, $c$, $b$, $t$,

$Y_\alpha = B$, if $\alpha$ is one of $d$, $u$.

Summing up the aforesaid we give a table of quantum characteristics of six quarks, which are a basis for constructing all known hadrons. 
\begin{table}[h] 
\caption{Table of quark characteristics} 
\label{tab_hadrons}
 \centering
\begin{tabular}{@{}c|c|c|c|c|c|c}
  \backslashbox{Characteristic}{Quark}& d& u& s& c& b& t
 \\\hline
  Electric charge $Q$ & $-\frac{1}{3}$ & $\frac{2}{3}$ & $-\frac{1}{3}$ & $\frac{2}{3}$ & $-\frac{1}{3}$ & $\frac{2}{3}$ 
 \\\hline
  Hypercharge $Y$ & $\frac{1}{3}$ & $\frac{1}{3}$ & $-\frac{2}{3}$ & $\frac{4}{3}$ & $-\frac{2}{3}$ & $\frac{4}{3}$
 \\\hline
  Baryon charge $B$ & $\frac{1}{3}$ & $\frac{1}{3}$ & $\frac{1}{3}$ & $\frac{1}{3}$ & $\frac{1}{3}$ & $\frac{1}{3}$
 \\\hline
  Spin $J$ & $\frac{1}{2}$ & $\frac{1}{2}$ & $\frac{1}{2}$ & $\frac{1}{2}$ & $\frac{1}{2}$ & $\frac{1}{2}$
 \\\hline
  Isospin $I$ & $\frac{1}{2}$ & $\frac{1}{2}$ & 0 & 0 & 0 & 0
 \\\hline
  Isospin projection $I_3$ & $-\frac{1}{2}$ & $\frac{1}{2}$ & 0 & 0 & 0 & 0
 \\\hline
  Flavour $d$ (down) & -1 & 0 & 0 & 0 & 0 & 0 
 \\\hline
  Flavour $u$ (up) & 0 & 1 & 0 & 0 & 0 & 0 
 \\\hline
  Flavour $s$ (strange) & 0 & 0 & -1 & 0 & 0 & 0 
 \\\hline
  Flavour $c$ (charm) & 0 & 0 & 0 & 1 & 0 & 0 
 \\\hline
  Flavour $b$ (beauty) & 0 & 0 & 0 & 0 & -1 & 0 
 \\\hline
  Flavour $t$ (true) & 0 & 0 & 0 & 0 & 0 & 1 
 \\\hline
\end{tabular}
\end{table}

\section*{Conclusion.}

In the present paper we have succeeded in establishing the relationship between the conserved quantum characteristics of elementary particles and the internal (hidden) symmetries of the six-dimensional pseudo-Euclidean space $\mathbb{E}_{3,3}$, which, as has been supposed, may be a real physical space of microcosm, limited by the time intervals of $10^{-20}$s. 
Proceeding from this concept, is turned possible to explain the phenomenon of violation of the conservation laws for hypercharge and isospin in weak interactions and keeping these laws in strong interactions, as well as to predict the number of possible quarks. 
It was shown that the conservation laws of the electric and baryon charges, hypercharge, spin and isospin are induced by the unitary symmetry group $SU(4)$, which represents the most general group of hidden symmetries of $\mathbb{E}_{3,3}$ space. 
The conservation laws of quantum characteristics such as color and flavour are generated by the hyperbolic groups of unitary symmetry $U(1,\mathbb{H})$ and $SU(2,\mathbb{H})$, which represent the symmetry groups of space $\mathbb{E}_{3,3}$ and are directly related to Lorentz group.

\pagebreak

\end{document}